\documentclass[reprint,aps,superscriptaddress]{revtex4-2}
\usepackage{float}
\usepackage{lineno}
\usepackage{upgreek}

\usepackage{graphicx}
\usepackage{dcolumn}
\usepackage{bm}

\usepackage{multirow}
\usepackage{booktabs}
\usepackage{makecell}
\usepackage{lineno}

\begin{document}


\title{Measurement of $\bm{e^+e^-}$ Momentum and Angular Distributions from Linearly Polarized Photon Collisions}


\affiliation{Abilene Christian University, Abilene, Texas   79699}
\affiliation{AGH University of Science and Technology, FPACS, Cracow 30-059, Poland}
\affiliation{Alikhanov Institute for Theoretical and Experimental Physics NRC "Kurchatov Institute", Moscow 117218, Russia}
\affiliation{Argonne National Laboratory, Argonne, Illinois 60439}
\affiliation{American University of Cairo, New Cairo 11835, New Cairo, Egypt}
\affiliation{Brookhaven National Laboratory, Upton, New York 11973}
\affiliation{University of California, Berkeley, California 94720}
\affiliation{University of California, Davis, California 95616}
\affiliation{University of California, Los Angeles, California 90095}
\affiliation{University of California, Riverside, California 92521}
\affiliation{Central China Normal University, Wuhan, Hubei 430079 }
\affiliation{University of Illinois at Chicago, Chicago, Illinois 60607}
\affiliation{Creighton University, Omaha, Nebraska 68178}
\affiliation{Czech Technical University in Prague, FNSPE, Prague 115 19, Czech Republic}
\affiliation{Technische Universit\"at Darmstadt, Darmstadt 64289, Germany}
\affiliation{ELTE E\"otv\"os Lor\'and University, Budapest, Hungary H-1117}
\affiliation{Frankfurt Institute for Advanced Studies FIAS, Frankfurt 60438, Germany}
\affiliation{Fudan University, Shanghai, 200433 }
\affiliation{University of Heidelberg, Heidelberg 69120, Germany }
\affiliation{University of Houston, Houston, Texas 77204}
\affiliation{Huzhou University, Huzhou, Zhejiang  313000}
\affiliation{Indian Institute of Science Education and Research (IISER), Berhampur 760010 , India}
\affiliation{Indian Institute of Science Education and Research (IISER) Tirupati, Tirupati 517507, India}
\affiliation{Indian Institute Technology, Patna, Bihar 801106, India}
\affiliation{Indiana University, Bloomington, Indiana 47408}
\affiliation{Institute of Modern Physics, Chinese Academy of Sciences, Lanzhou, Gansu 730000 }
\affiliation{University of Jammu, Jammu 180001, India}
\affiliation{Joint Institute for Nuclear Research, Dubna 141 980, Russia}
\affiliation{Kent State University, Kent, Ohio 44242}
\affiliation{University of Kentucky, Lexington, Kentucky 40506-0055}
\affiliation{Lawrence Berkeley National Laboratory, Berkeley, California 94720}
\affiliation{Lehigh University, Bethlehem, Pennsylvania 18015}
\affiliation{Max-Planck-Institut f\"ur Physik, Munich 80805, Germany}
\affiliation{Michigan State University, East Lansing, Michigan 48824}
\affiliation{National Research Nuclear University MEPhI, Moscow 115409, Russia}
\affiliation{National Institute of Science Education and Research, HBNI, Jatni 752050, India}
\affiliation{National Cheng Kung University, Tainan 70101 }
\affiliation{Nuclear Physics Institute of the CAS, Rez 250 68, Czech Republic}
\affiliation{Ohio State University, Columbus, Ohio 43210}
\affiliation{Panjab University, Chandigarh 160014, India}
\affiliation{Pennsylvania State University, University Park, Pennsylvania 16802}
\affiliation{NRC "Kurchatov Institute", Institute of High Energy Physics, Protvino 142281, Russia}
\affiliation{Purdue University, West Lafayette, Indiana 47907}
\affiliation{Rice University, Houston, Texas 77251}
\affiliation{Rutgers University, Piscataway, New Jersey 08854}
\affiliation{Universidade de S\~ao Paulo, S\~ao Paulo, Brazil 05314-970}
\affiliation{University of Science and Technology of China, Hefei, Anhui 230026}
\affiliation{Shandong University, Qingdao, Shandong 266237}
\affiliation{Shanghai Institute of Applied Physics, Chinese Academy of Sciences, Shanghai 201800}
\affiliation{Southern Connecticut State University, New Haven, Connecticut 06515}
\affiliation{State University of New York, Stony Brook, New York 11794}
\affiliation{Temple University, Philadelphia, Pennsylvania 19122}
\affiliation{Texas A\&M University, College Station, Texas 77843}
\affiliation{University of Texas, Austin, Texas 78712}
\affiliation{Tsinghua University, Beijing 100084}
\affiliation{University of Tsukuba, Tsukuba, Ibaraki 305-8571, Japan}
\affiliation{United States Naval Academy, Annapolis, Maryland 21402}
\affiliation{Valparaiso University, Valparaiso, Indiana 46383}
\affiliation{Variable Energy Cyclotron Centre, Kolkata 700064, India}
\affiliation{Warsaw University of Technology, Warsaw 00-661, Poland}
\affiliation{Wayne State University, Detroit, Michigan 48201}
\affiliation{Yale University, New Haven, Connecticut 06520}
\affiliation{Institute of Nuclear Physics PAN, Cracow 31-342, Poland}

\author{J.~Adam}\affiliation{Brookhaven National Laboratory, Upton, New York 11973}
\author{L.~Adamczyk} \affiliation{AGH University of Science and Technology, FPACS, Cracow 30-059, Poland}
\author{J.~R.~Adams}\affiliation{Ohio State University, Columbus, Ohio 43210}
\author{J.~K.~Adkins}\affiliation{University of Kentucky, Lexington, Kentucky 40506-0055}
\author{G.~Agakishiev}\affiliation{Joint Institute for Nuclear Research, Dubna 141 980, Russia}
\author{M.~M.~Aggarwal}\affiliation{Panjab University, Chandigarh 160014, India}
\author{Z.~Ahammed}\affiliation{Variable Energy Cyclotron Centre, Kolkata 700064, India}
\author{I.~Alekseev}\affiliation{Alikhanov Institute for Theoretical and Experimental Physics NRC "Kurchatov Institute", Moscow 117218, Russia}\affiliation{National Research Nuclear University MEPhI, Moscow 115409, Russia}
\author{D.~M.~Anderson}\affiliation{Texas A\&M University, College Station, Texas 77843}
\author{A.~Aparin}\affiliation{Joint Institute for Nuclear Research, Dubna 141 980, Russia}
% ...

\collaboration{STAR Collaboration}\noaffiliation

\date{\today}
\begin{abstract}

 The Breit-Wheeler process which produces matter and anti-matter from photon collisions is investigated experimentally through the observation of 6085 exclusive electron-positron pairs in ultra-peripheral Au+Au collisions at $\sqrt{s_{_{NN}}}=200$ GeV. 
 The measurements reveal a large fourth-order angular modulation of $\cos{4\Delta\phi}=(16.8\pm2.5)\%$ and smooth invariant mass distribution absent of vector mesons ($\phi$, $\omega$ and $\rho$) at the experimental limit of $\le 0.2\%$ of the observed yields. 
 The differential cross section as a function of $e^+e^-$ pair transverse momentum $P_\perp$ peaks at low value with $\sqrt{ \langle P_\perp^2 \rangle } = 38.1\pm0.9$ MeV and displays a significant centrality dependence. 
 These features are consistent with QED calculations for the collision of linearly polarized photons quantized from the extremely strong electromagnetic fields generated by the highly charged Au nuclei at ultra-relativistic speed.
 The experimental results have implications for vacuum birefringence and for mapping the magnetic field which is important for emergent QCD phenomena. 
\end{abstract}

\maketitle

When an electron at rest annihilates with its antimatter counterpart, a positron~\cite{anderson_apparent_1932}, the process results in the isotropic and monochromatic emission of two photons~\cite{Chao1930PhysRev.36.1519,klemperer_1934}. In 1934, Breit and Wheeler studied the theory of the reverse process of ``collision of two light quanta''~\cite{Breit-wheeler1934zz} to create electron-positron pairs. 
The original Breit-Wheeler study~\cite{Breit-wheeler1934zz} realized the near impossibility of achieving $\gamma$-ray collisions in existing Earth-based experiments and proposed an alternative approach with photon collisions originating from highly charged nuclei passing each other at ultra-relativistic speeds. 
Breit and Wheeler derived the cross section for photon-photon fusion ($\sigma_{\gamma\gamma}$) into $e^+e^-$ pairs, and used the work from Williams and Weizs{\"a}cker~\cite{weizsacker_ausstrahlung_1934,williamsNatureHighEnergy1934} demonstrating that a Lorentz-boosted Coulomb field in a certain kinematic phase space, propagated as a nearly transverse electromagnetic wave, can be quantized into a flux of real photons in the so-called equivalent photon approximation (EPA) to establish a viable source of photons. 

Since photons are spin 1 particles, in general their helicity ($J_z$) may take values $-1, 0,$ or $+1$. 
While real photons are massless and do not allow the $J_z=0$ state, short-lived virtual photons may carry a virtual mass (virtuality) with a possible $J_z=0$ state in their role as an intermediate propagator of the electromagnetic force. The consequences for the produced $e^+e^-$ in a collision of two real photons are a dramatic suppression of the production of vector mesons (spin 1 particles) and a preferential alignment of the $e^{\pm}$ momentum along the photon propagation axis (i. e., an anisotropic distribution in the polar
angle $\theta$).  

\begin{figure}
    \centering
    \includegraphics[width=0.43\textwidth]{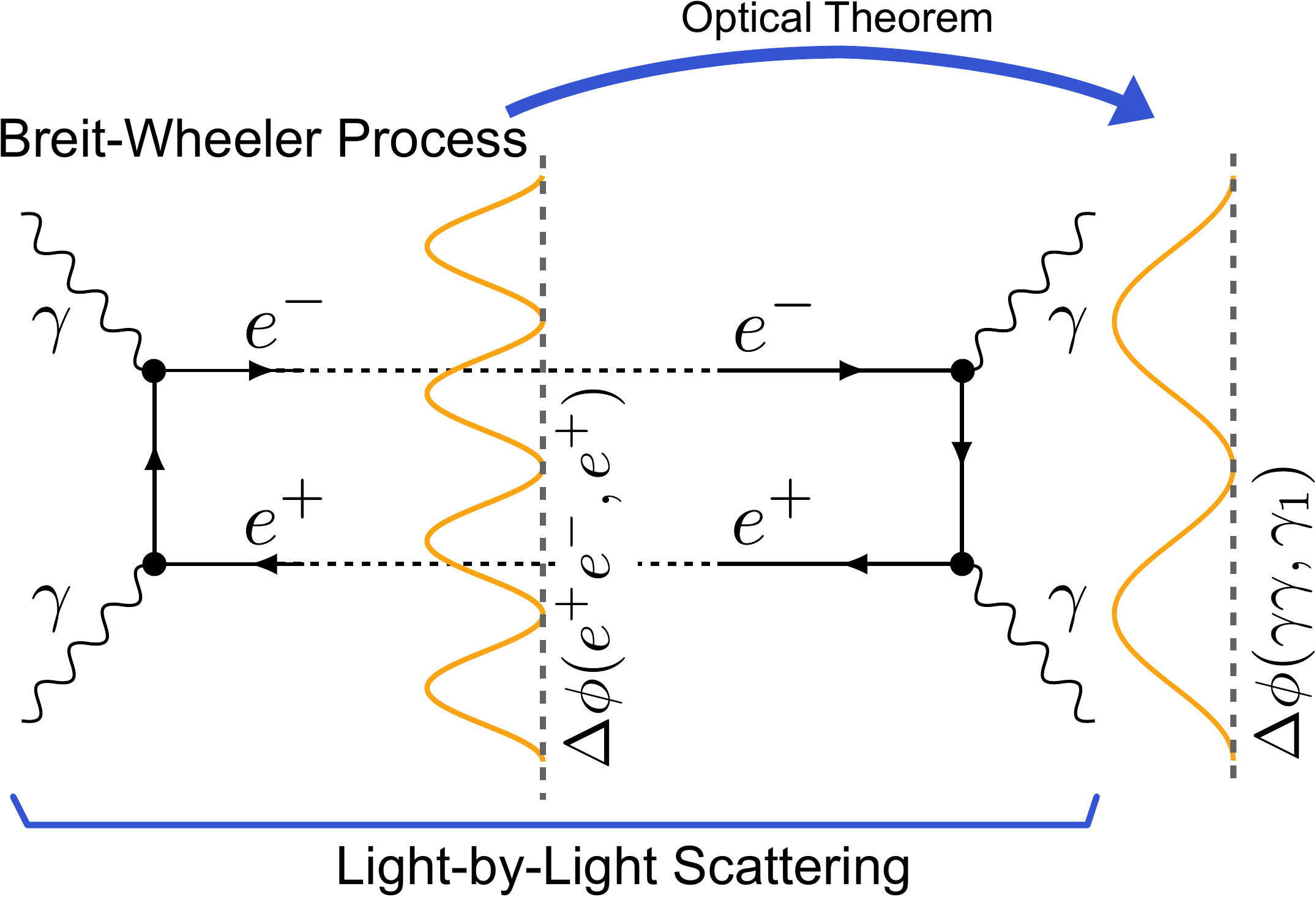}
    \caption{ A Feynman diagram for the exclusive Breit-Wheeler process and the related Light-by-Light scattering process illustrating the unique angular distribution predicted for each process due to the initial photon polarization.
    }
    \label{fig:fig1a}
\end{figure}

Another consequence of the quantum nature of the real photon intrinsic spin and wave-function is that the parallel and perpendicular relative polarization angles in photon-photon collisions result in distinct differential cross sections (Eq.~5.12 in Ref.~\cite{budnev_two-photon_1975} and Eq.~15 in Ref.~\cite{harland-lang_exclusive_2019}). 
It was realized only recently that these effects could be accessed experimentally in ultra-relativistic heavy-ion collisions~\cite{li_probing_2019} since the transverse momentum of the pair is correlated with the polarization of the photons. For linearly polarized photons, the distinct differential cross sections lead to a $\cos 4\Delta\phi$ angular distribution (see Fig.~\ref{fig:fig1a}), where $\Delta\phi$ is the azimuthal angle in the laboratory frame between the momentum of the $e^+e^-$ pair and one of the daughters  ($e^{\pm}$)~\cite{li_probing_2019}. 
Fundamentally, the angular modulations, both in $\theta$ and $\phi$, come about because the total spin of the $J=2$ composite state must be encoded into the orbital angular momentum of the daughter particles.

In Quantum Electrodynamics (QED), different processes of creating an $e^+e^-$ pair from the collision of two photons are defined depending on the virtuality of the photons and on whether the consideration of higher-order processes is necessary. There are three possible interactions: the collisions of two virtual photons (as calculated by Landau and Lifshitz, giving the total cross section for $e^+e^-$ production predominantly at the pair threshold~\cite{Landau1934}), of one virtual and one real photon (Bethe-Heitler process~\cite{bethe_h_stopping_1934}), or of two real photons --- the Breit-Wheeler process~\cite{Breit-wheeler1934zz}. It is important to note that all three processes can be identified in particle colliders in specific kinematics~\cite{PhysRevSTAB.17.051003,hencken_electromagnetic_1994}. The Breit-Wheeler process results in a strong constraint on the allowed energy distribution, and is only applicable for the production of $e^+e^-$ pairs at large angles (with respect to the beam axis) with large invariant mass at very small pair transverse momentum.

Creative experimental designs and increasing laser power may render the exclusive Breit-Wheeler process achievable at laser facilities~\cite{ruffini_electron-positron_2010,burke_positron_1997,ribeyre_pair_2016,Pike2014NaturePhotonics} in the near future.
While the production of $e^+e^-$ pairs via virtual photons is commonplace in high-energy collider experiments, it has also become well established in past decades that various processes involving real photons can be achieved by harnessing the photons of a highly Lorentz-contracted Coulomb field. 
As an example, we note the recent observations of Light-by-Light (LbyL) scattering achieved by harnessing the same source of photons~\cite{aaboud_evidence_2017,atlas_collaboration_observation_2019,sirunyan_evidence_2019}. 
The LbyL process is similar to the Breit-Wheeler process in that it involves the collision of two real photons in the initial state. 
According to the optical theorem, the Breit-Wheeler process and the $e^+e^-$ channel of LbyL scattering (see Fig~\ref{fig:fig1a}) are two parts of the same process --- the Breit-Wheeler process is the $\gamma\gamma$ absorption part and the LbyL scattering is the transmission part.
 
Over the decades, the production of $e^{+}e^{-}$ pairs has been studied at a wide array of hadron and $e^+e^-$ collider experiments~\cite{NA45BAUER1994471, star_collaboration_production_2004,LOW_ALICE,acciarri_production_1997,adeva_electroweak_1988,abbiendi_total_2000,vane_electron-positron_1992,afanasiev_photoproduction_2009,abbas_charmonium_2013,chatrchyan_search_2012,sirunyan_observation_2018,sirunyan_evidence_2019,aad_measurement_2015,aaboud_evidence_2017,WA93PhysRevLett.69.1911}. 
However, the existing experimental searches for the Breit-Wheeler process have not explored its unique features whereby the colliding photons have the energy spectrum and quantum spin states of real photons, and whereby any approximations do not alter the physics result of real photon collisions. 
Measurement of exclusive photon-mediated processes in ultra-relativistic nuclear collisions requires that the nuclei pass one another with an impact parameter ($b$) larger than the nuclear diameter, in so-called ultra-peripheral collisions (UPCs), such that no strong interactions can take place~\cite{baur_electron-positron_2007}. 
Measurement of the exclusive Breit-Wheeler process further requires a technique for isolating the collision of photons in UPCs. 

In this Letter, a comprehensive analysis includes simultaneous measurement of
(a) the total $\gamma\gamma\rightarrow e^+e^-$ production rate, (b) the photon energy spectrum with sufficient precision to demonstrate the relationship with the initial spatial distribution of the electromagnetic field, and (c) the allowed helicity states for participating photons via measurement of the polar angle of the produced positrons and via measurement of the invariant mass spectra to demonstrate the absence of vector mesons.
Furthermore, we present the first measurement of the unique $\cos{4\Delta\phi}$ modulation predicted for the Breit-Wheeler photon-photon fusion process to definitively demonstrate that the interacting photons behave as real photons with transverse linear polarization. 


\begin{figure}
    \centering
    \includegraphics[width=0.43\textwidth]{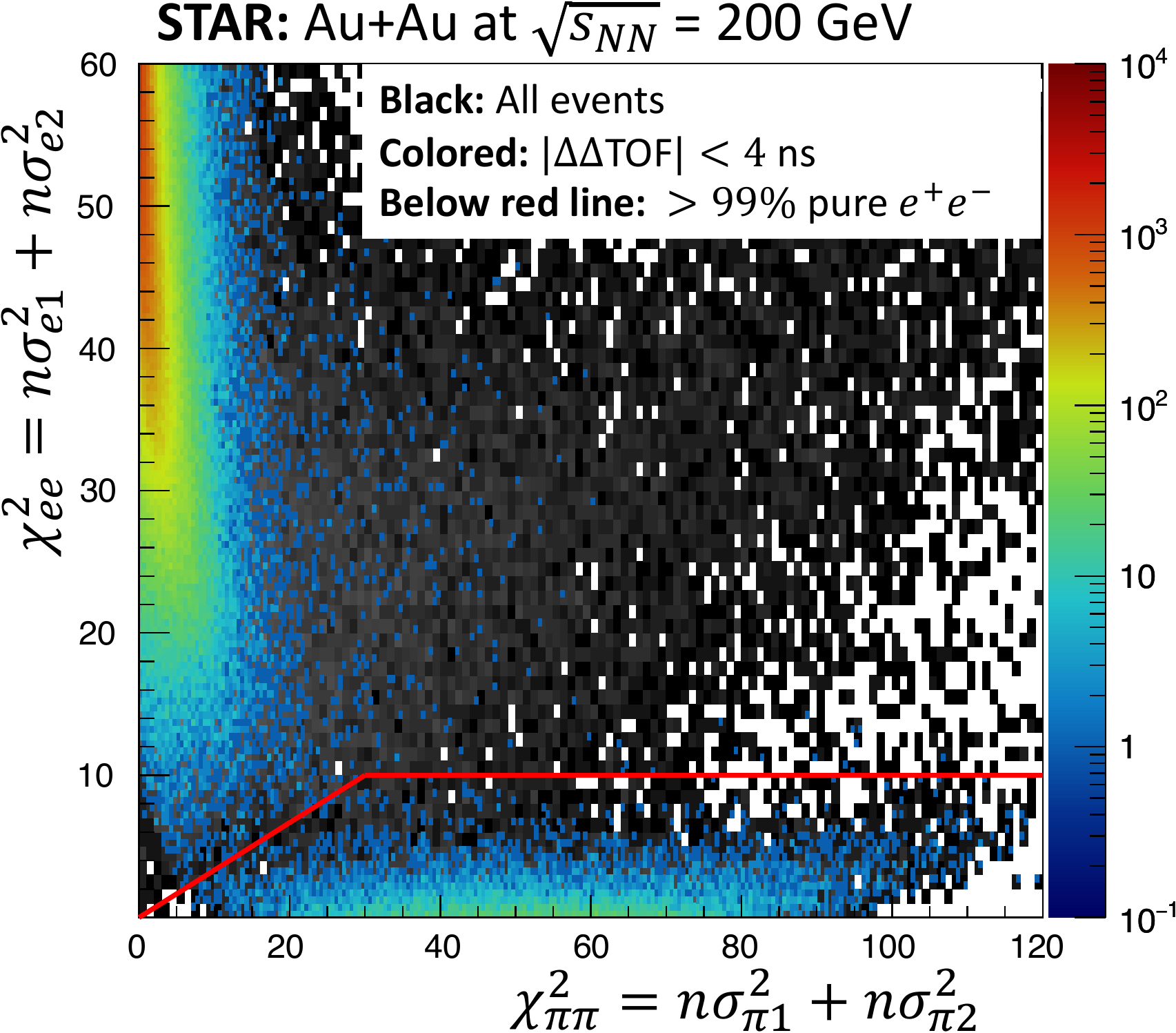}
    \caption{The $\chi^2_{ee}$ vs.  $\chi^2_{\pi\pi}$ distribution 
    before (after) applying the $|\Delta\Delta$ TOF$|$ $<0.4$ ns criteria in black (color). The $e^+e^-$ candidates are shown in color below the red lines. 
    }
    \label{fig:fig1b}
\end{figure}

\begin{figure*}
    \centering
    \includegraphics[width=1\textwidth]{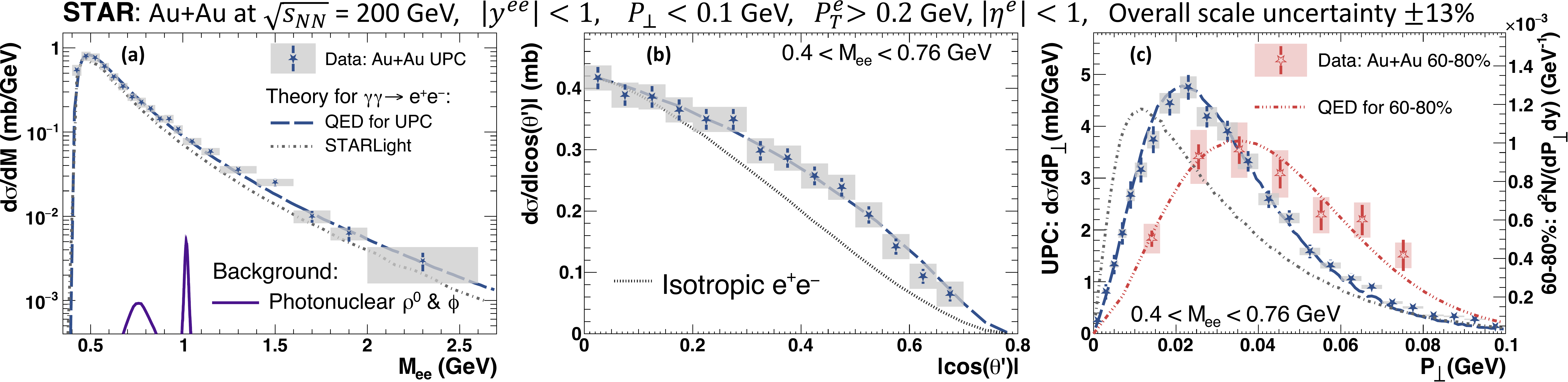}
    \caption{ The fully corrected differential cross sections for exclusively produced $e^+e^-$ pairs with respect to (a) the invariant mass $M_{ee}$ (and predicted vector-meson background from photoproduction~\cite{klein_starlight_2017}), (b)  the polar angle distribution $|\cos\theta^\prime|$, and (c) the pair transverse momentum $P_\perp$.
    }
    \label{fig:fig2}
\end{figure*}

This measurement of exclusive $e^+e^-$ pair production was conducted at the Relativistic Heavy Ion Collider (RHIC) by the Solenoidal Tracker at RHIC (STAR) Collaboration. The measurement uses gold-gold (Au+Au) collisions at a center of mass energy per nucleon pair ($\sqrt{s_{_{NN}}}$) of 200 GeV. A triggering system based on signals from several STAR detectors is used to select UPC events that may contain exclusive $e^+e^-$ pairs~\cite{judd_evolution_2018}, in conjunction with the excitation and dissociation of the passing gold nuclei. This tagging process of mutual Coulomb dissociation~\cite{bertulani_physics_2005,baltz_two-photon_2009} at RHIC is well modeled~\cite{Baltz:1998ex} with a cross section uncertainty of $\pm5\%$~\cite{ZDC130xs2002,supp}.
In total, 23$\times10^6$ events were analyzed from the UPC triggered data taken in the year 2010. The recorded dataset corresponds to an integrated luminosity of $\int\mathcal{L}dt=700\pm70~\upmu\mathrm{b}^{-1}$. 

The most common background process meeting our trigger requirement is the photo-nuclear production of a $\rho^0$~meson which decays to $\pi^+\pi^-$~\cite{Adamczyk:2017vfu}. 
For this reason, high purity identification of $e^+e^-$ pairs is a crucial ingredient in this measurement. 
Clean $e^+e^-$ pairs are identified using the measured ionization energy loss by constructing a $\chi^2_{ee(\pi\pi)} = n\sigma_{e1(\pi1)}^2 + n\sigma_{e2(\pi2)}^2$ value, where $n\sigma_{e(\pi)}$ is the number of standard deviations from the expectation for an electron or pion mass hypothesis, respectively.
Contamination from hadron pairs is further reduced using the double difference in the time-of-flight ($\Delta\Delta$TOF) between the two measured tracks and the expectation for an $e^+e^-$ pair calculated using the measured track momentum and path length~\cite{supp}.
Figure~\ref{fig:fig1b} shows the distribution of candidate pairs before (in black) and after (in color) applying the $\Delta\Delta$TOF requirement.
Together these selection criteria achieve better than 99\% pure $e^+e^-$ selection.

In addition to the measurements in UPCs, we also present measurements from collisions in $60-80\%$ centrality in which the nuclei interact via the strong force with an impact parameter between approximately $11.5$ and $13.5$ fm~\cite{PhysRevLett.121.132301}. For these events, the hadronic and medium-induced background in the selected kinematic range is at the level of a few percent and is subtracted statistically. For more details on the selection and analysis of these events, see Ref.~\cite{PhysRevLett.121.132301}.
\begin{table}[t]
\centering
\begin{tabular}{@{}l l l l l l l@{}} 
\textbf{Quantity} & \multicolumn{3}{c}{ \textbf{Measured} } & \textbf{SL} & \textbf{gEPA} & \textbf{QED} \\ \toprule
${\sigma}$ $(\upmu\mathrm{b})$ & \multicolumn{3}{c}{ \shortstack{ 261$\pm$4$\pm$13$\pm$34} } & 220 & 260 & 260 \\ \midrule
& \multicolumn{4}{c}{ \textbf{Ultra-Peripheral}} & \multicolumn{2}{c}{ \textbf{Peripheral}} \\ \cmidrule(lr){2-5} \cmidrule(lr){6-7}
                                                           & Measured                      & QED     & SC & SL  & Measured           & QED      \\ \toprule
${|A_{4\Delta\phi}|}$ (\%)                                   & $16.8 \pm 2.5$                & 16.5    & 19 & 0   & $27\pm6$           & 34.5     \\ \midrule
${|A_{2\Delta\phi}|}$ (\%)                                   & $2.0\pm2.4$                   & 0       & 5 & 5   & $6\pm6$            & 0        \\ \midrule
$\sqrt{ \langle P_\perp^2 \rangle }$ (MeV) & $38.1\pm0.9$ & 37.6 & 35.4 & 35.9 & $50.9\pm2.5$ & 48.5 \\ \bottomrule

\end{tabular}
\caption{ \label{tab:results} 
Top row: cross section within the fiducial STAR acceptance~\cite{supp} for $\gamma\gamma\rightarrow e^+e^-$ compared with theory calculations~\cite{li_impact_2020,zha_initial_2020,klein_starlight_2017} (SL stands for STARLight, SC for SuperChic). The quoted uncertainties on the measured cross section are for statistical, systematic and the overall scale uncertainty, respectively.
Lower rows: $\Delta\phi$ and $\sqrt{\langle P_\perp^2 \rangle}$ from UPCs and $60-80\%$ central collisions (peripheral) with the corresponding theory calculations~\cite{li_impact_2020,zha_initial_2020,harland-lang_exclusive_2019,klein_starlight_2017} where applicable. The fits to the data with Eq.~(\ref{eq:fs}) result in $\chi^2/\mathrm{ndf}$ of $19/16$ and $10/17$ for UPC and $60-80$\% centrality, respectively. 
The quoted uncertainties are statistical and systematic uncertainties added in quadrature. }
\end{table}
The cross section for exclusively produced $e^+e^-$ pairs was measured in a fiducial phase space defined by the acceptance for daughter particles, corresponding to pairs with an invariant mass of $0.4 < M_{ee} < 2.6$ GeV and with transverse momentum of $P_\perp < 0.1$ GeV. 
The measured fiducial cross section is 261 $\pm$4~(stat.) $\pm$13~(syst.) $\pm$34~(scale uncertainty~\cite{supp}) $\upmu$b for events with one or more neutrons emitted in each beam direction. 
Measurements of the production rate for exclusive $e^+e^-$ pairs, fully corrected for event selection and detector effects, are shown in the three panels of Fig.~\ref{fig:fig2}. All observables are reported for kinematic acceptance within $P_\perp<0.1$ GeV with the $M_{ee}$ limits noted in each panel.

Figure~\ref{fig:fig2}(a) shows the invariant mass of exclusive $e^+e^-$ pairs. The invariant mass spectrum is smooth and featureless even in the range of known vector mesons~\cite{PDG2018PhysRevD.98.030001}. This is a consequence of the quantum numbers of the two photons involved in the Breit-Wheeler process~\cite{brodsky_large-angle_1981} where helicity state $J_z = 0$ is absent for real photons but necessary for exclusive single vector-meson production. 
Fits to the Breit-Wheeler shape plus the vector meson's mass spectral line shape show the absence of all light vector mesons and result in the following limits to the measured $e^+e^-$ cross section: $\rho$ at $(-0.4\pm1.2)\%$, $\omega$ at $(-0.5\pm0.3)\%$ and $\phi$ at $(0.2\pm0.2)\%$. Potential background contribution from exclusive photo-nuclear production of vector mesons~\cite{star_collaboration_coherent_2017} with the decay branch $\rho^0 (\phi)\rightarrow e^+e^-$ is simulated in STARLight and shown as purple lines in Fig.~\ref{fig:fig2}(a).  
The STARLight model is also used to predict the background from double vector meson production (e.g. $\gamma\gamma\rightarrow\rho^0\rho^0$) where one vector meson decays to an $e^+e^-$ pair.
The cross section for such a process is several orders of magnitude lower than the exclusive photoproduction of a single $\rho^0$~\cite{star_collaboration_coherent_2017}. In addition, such a process of semi-inclusive $\rho^0$ production results in a broad $\rho^0$ transverse momentum distribution and is estimated to be less than $10^{-5}$ times the already negligible background contribution from photo-nuclear production of $\rho^0$. 

Figure~\ref{fig:fig2}(b) shows the $|\cos\theta^\prime|$ distribution, in which $\theta^\prime$ is the polar angle of the $e^+$ momentum vector with respect to the beam, measured in the $e^+e^-$ center-of-mass frame. 
The main structure, the fall-off of $|\cos\theta^\prime|$, is the result of the gross detector acceptance that limits detection of particles to $45^\circ\lesssim\theta\lesssim 135^\circ$. 
However, the Breit-Wheeler process exhibits an enhancement toward small polar angle, measurably different from that of isotropic $e^+e^-$ emission. 
The contribution from isotropic $e^+e^-$ emission is determined via a template fit and found to be consistent with zero [$(1\pm2)$\% of the measured cross section].

In Fig.~\ref{fig:fig2}(c), we show the differential cross section as a function of the pair transverse momentum ($P_\perp$) in UPCs compared with the same distribution in $60-80\%$ central collisions to demonstrate the sensitivity of the process to the initial geometry of the colliding electromagnetic fields. The data show a clear peak in the production rate at very low $P_\perp$. 
The shapes of the spectra are quantified by the spread in the transverse momentum plane (via $\sqrt{\langle P_\perp^2 \rangle}$) calculated from the data where available plus an exponential fit to estimate the additional contribution above the measured range~\cite{PhysRevLett.121.132301} (see Table~\ref{tab:results}). 

\begin{figure}
    \centering
    \includegraphics[width=0.45\textwidth]{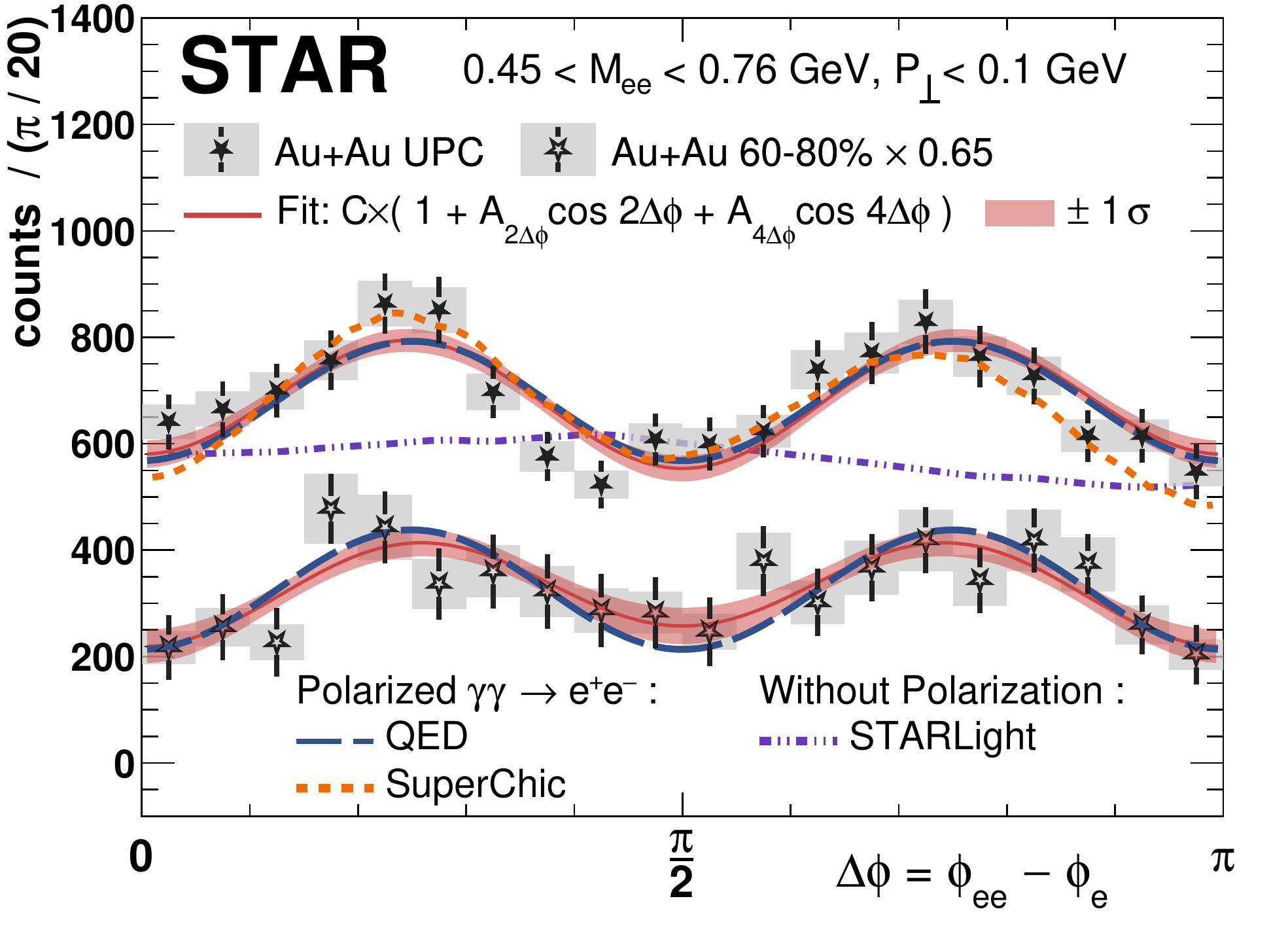}
    \caption{ The $\Delta\phi = \phi_{ee} - \phi_e$ distribution from UPCs and $60-80\%$ central collisions for $M_{ee}>0.45$ GeV with calculations from  QED~\cite{li_impact_2020}, STARLight~\cite{klein_starlight_2017} and from the publicly available SuperChic3 code~\cite{harland-lang_exclusive_2019}. }
    \label{fig:fig3}
\end{figure}

Finally, Fig.~\ref{fig:fig3} shows the first measurement of the angular distribution $\Delta\phi$ for $e^+e^-$ pairs produced in photon-photon collisions. Distributions from both UPCs and $60-80\%$ central collisions are shown with fits to a function of the form:

\begin{equation}
    f(\Delta\phi) = C ( 1 + A_{2\Delta\phi} \cos{ 2\Delta\phi }  + A_{4\Delta\phi} \cos{ 4\Delta\phi } ),
    \label{eq:fs}
\end{equation}

\noindent where $C$ is a constant and $A_{2\Delta\phi}$ ($A_{4\Delta\phi}$) is the magnitude of a $\cos{ 2\Delta\phi }$ ($\cos{ 4\Delta\phi }$) modulation. 
The observed magnitude of the $\cos{ 2\Delta\phi }$ and $\cos{ 4\Delta\phi }$ modulations are reported in Table~\ref{tab:results}. 
These data were not unfolded to remove momentum resolution effects, which contribute a +1.5\% and +3.5\% correction for UPCs and $60-80\%$ central collisions, respectively~\cite{supp}. The data presented in Figs.~\ref{fig:fig2} and \ref{fig:fig3} are plotted with statistical (vertical bars) and systematic (boxes) uncertainties~\cite{supp}. 

The measured fiducial cross section is compared with two calculations that incorporate mutual Coulomb excitation, nuclear dissociation, and the production of $e^+e^-$ pairs according to the Breit-Wheeler photon-photon fusion cross section. The QED theory is a numerical calculation of the differential cross sections at the lowest-order QED as illustrated in Fig.~\ref{fig:fig1a}. The prescription in Ref.~\cite{hencken_electromagnetic_1994} was followed in a new implementation in Ref.~\cite{zha_initial_2020}. The STARLight model~\cite{klein_starlight_2017} implements a conventional EPA, factorizes photon flux into energy and transverse momentum spectra independently and excludes the photon flux inside nuclei. The consequential features are a lower cross section due to the exclusion as shown in Fig.~\ref{fig:fig2}(a), a softer $P_{\perp}$ distribution independent of impact parameter as shown in Fig.~\ref{fig:fig2}(c) and the absence of any azimuthal anisotropy. We list the predicted total cross section within the STAR acceptance from these calculations (Table~\ref{tab:results}). A third model calculation using generalized EPA (gEPA) is also presented. It performs a multi-dimensional integration of the form factors and the Breit-Wheeler cross section over the specific impact parameter~\cite{zha_initial_2020}. The total measured cross section agrees with all three calculations at the $\pm 1 \sigma$ level.  
The distributions presented in Figs. \ref{fig:fig2} and \ref{fig:fig3} are all, within uncertainties, consistent with the expectation from the Breit-Wheeler process alone. We observe a significant ($4.8\sigma$) increase in the $\sqrt{\langle P_\perp^2 \rangle}$ in $60-80\%$ central collisions compared to the same quantity in UPCs. 
For the $60-80\%$ central data, the large uncertainties allow room for some additional broadening of the $P_\perp$ distribution. 
A best fit value is found using the Breit-Wheeler distribution convoluted with a Gaussian having a width of $\sigma = 14 \pm 4{\rm (stat.)} \pm 4{\rm (syst.)}$~MeV ($\chi^2/$ndf $=3.4/6$). 
These data demonstrate that the energy spectrum of the colliding photons depends on the nucleus-nucleus impact parameter, and therefore, on the spatial distribution of the electromagnetic fields. 
Both spectra are well described (total production rate and differential shape) by the QED calculations which include this dependence~\cite{zha_initial_2020,li_impact_2020} and invalidate several existing models~\cite{klein_starlight_2017,zha_initial_2020,harland-lang_exclusive_2019,li_probing_2019} that neglect it. These observed features of the Breit-Wheeler process provide experimental confirmation of fundamental QED predictions.

 
In UPCs, the $\cos{ 4\Delta\phi }$ modulation is observed with an amplitude of ($16.8\pm2.5$)\%. The data are in good agreement with numerical lowest-order QED calculations which predict an amplitude of $16.5$\%.
The data are also compared to predictions from the STARLight~\cite{klein_starlight_2017} and SuperChic~\cite{harland-lang_exclusive_2019} models.
STARLight, which includes the single-photon kinematics for the process but does not employ any polarization-dependent effects, predicts an isotropic distribution. 
SuperChic is a model similar to STARLight, but with the photon helicity dependence determined by the orientation of the electromagnetic fields in the transverse plane. 

When the collisions are defined as a flux of photons from the projectile nucleus traversing a circular magnetic field generated by the target nucleus~\cite{Hattori:2012je,hattori_photon_2016,kharzeev_chiral_2016,BFieldKharzeev:2007jp}, the observation of a separation in the differential angular distribution of the produced particles relative to the initial photon polarization and magnetic field angle is closely related to the phenomenon of birefringence. 
The striking thing about this observation is that it occurs through the electromagnetic field which polarizes the vacuum in the absence of a medium~\cite{heisenberg_consequences_2006}.

Vacuum birefringence~\cite{heisenberg_consequences_2006} is a phenomenon in which the refractive index of empty space depends on the relative angle between the photon polarization and the magnetic field direction. The only other evidence associated with vacuum birefringence since its prediction has been the observation of an enhanced linear polarization of light reaching Earth after traversing strong magnetic fields generated by pulsars~\cite{Mignani:2016fwz}.
Since photons, and more generally electromagnetic fields, cannot interact with each other directly, the Feymann diagram shown in Fig.~\ref{fig:fig1a} represents the simplest process by which such an interaction can occur, and relates to the imaginary part of the refractive index~\cite{Hattori:2012je,hattori_photon_2016}. Some authors suggested that such a phenomenon is more closely related to the vacuum dichroism~\cite{hattoriDileptonProductionSingle2021}. However, direct measurement of the real part of the refractive index of the vacuum birefringence effect through a similar azimuthal angular distribution of outgoing photons in UPCs, as proposed in Ref.~\cite{hattori_photon_2016}, may be quite challenging. Similar proposed experiments that make use of ultra-strong lasers and their generated standing electromagnetic fields are likely possible in the not-so-distant future~\cite{bragin_high-energy_2017}.  

Since the photons are linearly polarized along the radial direction around the nuclei, only a probe that is sensitive to the spatial and momentum distribution of the field (commonly known as the Wigner function) can utilize the polarization information. Therefore, our measurements of the energy spectrum and angular distribution provide the information needed to map the spatial extent of the intense electromagnetic fields produced by ultra-relativistic heavy nuclei for the first time and can be compared to different models which incorporate the charge distribution~\cite{Shou:2014eya} with a Lorentz boost. It has been proposed that fluctuations in the nucleon distribution inside a large nucleus can generate a much larger magnetic field strength with random orientation near the center of the nucleus~\cite{Skokov:2009qp}. Such a field configuration would result in a larger Breit-Wheeler cross section in the high $P_{\perp}$ tail and a reduced $\cos{4\Delta\phi}$ toward central collisions. 
The assumptions in such models are different from the QED calculations and models in comparison here that employ a uniform and continuous charge distribution inside the nucleus. 
As an example of this constraining power, we performed a fit of the QED calculations to our measured $d\sigma/dP_\perp$ distribution, assuming a continuous Woods-Saxon charge distribution. We observe a best fit ($\chi^2/{\rm dof}=8.0/9$) for $R=6.7\pm0.2$ fm and $a=0.2\pm0.2$ fm (though we note that these parameters are highly anti-correlated~\cite{supp}).
These observations and future work supply novel input about the electromagnetic fields that may drive yet undiscovered magnetohydrodynamical phenomena of QCD.  
\section*{Acknowledgments}
We thank the RHIC Operations Group and RCF at BNL, the NERSC Center at LBNL, and the Open Science Grid consortium for providing resources and support.  This work was supported in part by the Office of Nuclear Physics within the U.S. DOE Office of Science, the U.S. National Science Foundation, the Ministry of Education and Science of the Russian Federation, National Natural Science Foundation of China, Chinese Academy of Science, the Ministry of Science and Technology of China and the Chinese Ministry of Education, the National Research Foundation of Korea, Czech Science Foundation and Ministry of Education, Youth and Sports of the Czech Republic, Hungarian National Research, Development and Innovation Office, New National Excellency Programme of the Hungarian Ministry of Human Capacities, Department of Atomic Energy and Department of Science and Technology of the Government of India, the National Science Centre of Poland, the Ministry  of Science, Education and Sports of the Republic of Croatia, RosAtom of Russia and German Bundesministerium fur Bildung, Wissenschaft, Forschung and Technologie (BMBF) and the Helmholtz Association.

\bibliography{biblio}

\end{document}